\documentclass[graybox]{svmult}

\usepackage{mathptmx}       
\usepackage{helvet}         
\usepackage{courier}        
\usepackage{type1cm}        
\usepackage{makeidx}         
\usepackage{graphicx}        
\usepackage{multicol}        
\usepackage[bottom]{footmisc}
\usepackage{cite}

\makeindex             

\begin{document}

\title*{High Performance Calculation of Magnetic Properties and Simulation of Nonequilibrium Phenomena in nanofilms}
\titlerunning{HPC of Magnetic Properties and Nonequilibrium Phenomena in nanofilms}
\author{V.Yu. Kapitan and K.V. Nefedev}
\institute{V.Yu. Kapitan \at Far Eastern Federal University, The School of Natural Sciences, 8 Sukhanova St., Vladivostok 690950, Russia, \email{kvy@live.ru}
\and K.V. Nefedev \at Far Eastern Federal University, The School of Natural Sciences, 8 Sukhanova St., Vladivostok 690950, Russia, \email{knefedev@phys.dvgu.ru}}
%
%
\maketitle

\abstract*{Images of surface topography of ultrathin magnetic films have been used for Monte Carlo simulations in the framework of the ferromagnetic Ising model to study the hysteresis and thermal properties of nanomaterials. For high performance calculations was used super-scalable parallel algorithm for the finding of the equilibrium configuration. The changing of a distribution of spins on the surface during the reversal of the magnetization and the dynamics of nanodomain structure of thin magnetic films under the influence of changing external magnetic field was investigated.}

\abstract{Images of surface topography of ultrathin magnetic films have been used for Monte Carlo simulations in the framework of the ferromagnetic Ising model to study the hysteresis and thermal properties of nanomaterials. For high performance calculations was used super-scalable parallel algorithm for the finding of the equilibrium configuration. The changing of a distribution of spins on the surface during the reversal of the magnetization and the dynamics of nanodomain structure of thin magnetic films under the influence of changing external magnetic field was investigated.}

\section{Introduction}
\label{sec:1}

The need for theoretical research and a simulation of the physical properties of ultrathin ferromagnetic films due to the existence of fundamental problems of physics of magnetic phenomena, as well as the need for development of the theory of ferromagnetism, in general, and a hysteresis phenomena in particular. Computer processing of experimental data and subsequent simulation of the behavior of the surface of the magnet on the basis of these data allows obtain a new information about the physical nature of ferromagnetism and ferromagnetic anisotropy, visualize the processes of reversal of the magnetization in external fields. The research of the physical properties of thin ferromagnetic films is need from point of view of their practical applications in microelectronics and computer technology, as nanostructured soft magnetic thin magnetic films are currently the main materials for the manufacture of components of the magnetic random access memory (magnetic random access memory (MRAM)) ~\cite{science-journal1,science-journal2,
science-journal3,science-journal4,
science-journal5,science-journal6}.

The development of numerical and supercomputing methods provides new classes of algorithms that can solve complex problems of numerical simulation, handle large and superlarge volumes of data. Moreover, the level of sampling elements in a model of today is determined by resolution of the scanning tunneling microscope or atomic force microscope. The aim of this work is to develop a computer model and the creation of application software for processing data obtained with a scanning tunneling microscope (STM) and atomic force microscope (AFM), as well as the calculation of the magnetic and structural properties of the quasi-nanocluster magnets and simulation of magnetic hysteresis phenomena.

\section{The Model}
\label{sec:2}

A method for obtaining samples and experimental data were published in ~\cite{science-journal7,science-journal8}. The essence of the proposed method of computer image processing and subsequent Monte Carlo (MC) simulations, is based on the processing of the raster STM and AFM images, that allows construct 3D space lattice of spins with coordination number equal 12. The brightness of a pixel in the STM image is a function of the distance between the tip and the surface, so the image pixels were used to construct a magnet with a given number of atomic layers, the number of which was controlled by experimental methods. The used algorithm is describes in detail in ~\cite{science-journal7}.

In the lattice sites located model elements - spins Si, whose values have changed abruptly from -1 upto +1.  In principle, the simulation can occur within any known model, such as the Ising model or the Heisenberg exchange integral a certain value. We have used the Ising model, where each spin in lattice model of nanofilm interact via direct exchange with its nearest neighbors (up to 12 neighbors). The Boltzmann constant $k = 1$ and the exchange integral $J = 1$ values where used in the Metropolis algorithm. The transition from these values to measured experimentally a physical quantities can be carried out in accordance with the expression ~(\ref{01}):

\begin{equation}
\label{01}
J=\frac{3kT_{c}}{2zS}(S+1)
\end{equation}
where $z$ - the number of nearest neighbors, $T_{c}$ - the Curie temperature, $S$- the spin of the ion.


\section{A parallel Metropolis algorithm}
\label{sec:3}

The algorithm of the processing of STM images is discussed in paper ~\cite{science-journal8}. The number of rows of three-dimensional array equals the height of the STM-image, and the number of columns, the width of the picture, besides the "depth" is set equal to the number of layers of Co in the sample (based on experimental data).

\begin{figure}[h]

\includegraphics[width=120mm]{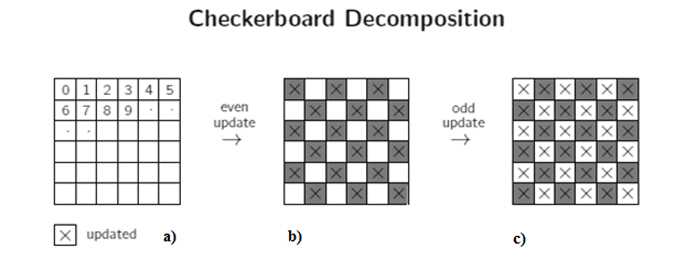}

\caption{Distribution of the matrix on the processes in a checkerboard pattern.}
\label{fig:1}       
\end{figure}

\begin{figure}[h]
\centering{\includegraphics[width=100mm]{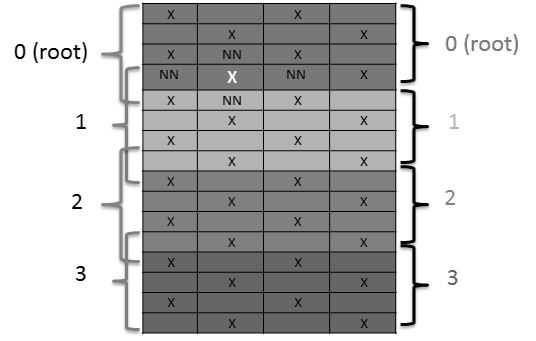}}
\caption{Distribution of the rows of blocks to process.}
\label{fig:2}       
\end{figure}

The parallelism of the algorithm is implemented by splitting the three-dimensional array of spins on the parts  (planes), for their subsequent distribution used the MPI library, and, accordingly, processing the each from a planes in the separate computation
process. Pass (MC step) on the spin system and the implementation of Monte Carlo (MC) steps in the algorithm of Metropolis produced in the "checkerboard decomposition". This is done to avoid the problem of boundary conditions calculated for the configurations of planes handled, see Fig.~\ref{fig:1}. The initial values did not change during the MC, i.e. in this case, for each step of temperature or field, half of MC steps is initially for one half of the spins Fig.~\ref{fig:1}(b), and then for the second half Fig.~\ref{fig:1}(c).
\\Parallel Metropolis algorithm is as follows:
\begin{enumerate}
\item The number of rows (planes) of three-dimensional array of spins was send in the each process. This number was proportional to the number (braces on the right in Fig.~\ref{fig:2}). We used the approach, in which the maximum number of processes for the execution of the program equals to one form two of the linear dimensions of the STM (AFM) image in pixels;
\item Since each spin of the cobalt samples, see [7], the highest possible number of nearest neighbors $z=12$, then for the correct accounting for spin neighbors standing on a block boundary rows (Fig.~\ref{fig:2}, marked by a white cross spin) the additional boundary rows of the neighboring processes sent in each process, which ones are located in different processes (braces on the left in Fig.~\ref{fig:2});
\item  The distribution of blocks of rows by using in-line functions $MPI\_Scatterv$, and the assembly, respectively, with the help of $MPI\_Gatherv$.
\end{enumerate}

The search of the equilibrium configuration for each part is performed using usual Monte Carlo (Metropolis algorithm). Within the framework the model hysteresis magnetic phenomena were explained as the effect of nonequilibrium in the Ising spin lattice model. For the simulation of nonequilibrium processes in the monolayer samples were used only surface points of model sample, and used the number of MC steps used proportionally increased with the number of lattice sites. For this reason, the system of correlated spins cannot go into a state of equilibrium during the time of field changing altering, which in our case leads to the phenomenon of magnetic hysteresis in the used model. The absence of an exact match of simulated and experimental data is due to the simplicity of the model.

For systems of a large number of Ising spins $N$, movement toward equilibrium may be slow, especially at low temperatures $T$. Therefore, in order to speed up getting the most probable configuration with a given energy and spin excess was used in parallel computing scheme. The scalability of the algorithms is provides by independent calculations for each row of the matrix, i.e. for each pixel rows in the STM (AFM) images and, consequently, maximum scalability is determined by the pixel resolution of the image. The splitting into two groups of processes using the $MPI$: $MPI\_Comm\_group$, can significantly increase the efficiency of the computation.

\section{The critical concentration and ferromagnetism}
\label{sec:4}

In ~\cite{science-journal9}, authors have presented a method to calculate the critical concentration, which one required for a phase transition to ferromagnetism in the crystal lattices with different numbers of nearest neighbors. We have determined the critical concentration of magnetic atoms for the transition to the ferromagnetic state for the monolayer and submonolayer samples of experimental data for which are given in ~\cite{science-journal7,science-journal8}. The critical concentration $p_{c}$ the transition to the ferromagnetic state at $T = 0$ were determined from the relation ~(\ref{02}):
\begin{equation}
\label{02}
p_{c}=\frac{2}{z}
\end{equation}

\begin{figure}[h]

\centering{\includegraphics[width=60mm]{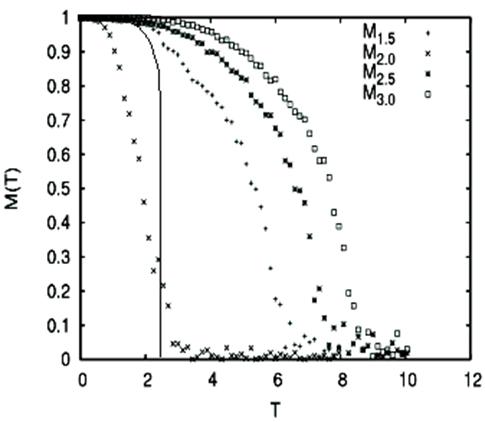}}

\caption{The temperature behavior of magnetization.}
\label{fig:3}       
\end{figure}

\begin{table}
\caption{Calculated concentrations of atoms in comparison with the critical concentrations and Curie temperature for different samples.}
\label{tab:1}       
%
%
\begin{tabular}{p{3.7cm}p{2.8cm}p{1.6cm}p{1.5cm}p{1.5cm}}
\hline\noalign{\smallskip}
The number of monolayers & The number of nearest neighbors & \  \ $T_{c}$ & $p, at\%$ & $p_{c}$ \\
\noalign{\smallskip}\svhline\noalign{\smallskip}
1.5 ML & 3.615  & 3.00 & 0.38 & 0.55\\
2.0 ML & 6.984 & 6.15 & 0.50 & 0.29\\
2.5 ML & 8.177 & 7.80 & 0.63 & 0.24\\
3.0 ML & 9.308  & 8.70 & 0.76 & 0.21\\
\noalign{\smallskip}\hline\noalign{\smallskip}
\end{tabular}

\end{table}
Table~\ref{tab:1} shows the critical concentration required for a phase transition to the ferromagnetic state, which implies that the sample of 1.5 ML at low temperatures should be in a state of the cluster ferromagnetism as $p<p_{c}$. That is confirmed by the data on the temperature dependence of magnetization for samples at 1.5, 2.0, 2.5. and 3.0 monolayer, shown in Fig.~(\ref{fig:3}).
\section{The critical field of switching}
\label{sec:5}
In the Ising model Hamiltonian $\mathcal{H}$ Ising spin system interacting with a direct exchange, in an external magnetic field~(\ref{03})
\begin{equation}
\label{03}
\mathcal{H}=-J\sum \limits_{n}^{} S_{n}S_{n+n_{0}}-h\sum \limits_{n}^{} S_{n}
\end{equation}
$n_{0}$ - number vector of the nearest neighbors of the crystalline host $n$. The partition function~(\ref{04})
\begin{equation}
\label{04}
Z_{N}(h,T)=\sum \limits_{s_{1}}^{}
\sum \limits_{s_{2}}^{}\cdot\cdot\cdot
\sum \limits_{s_{n}}^{}\cdot\cdot\cdot
\sum \limits_{s_{N}}^{}e^{[-\sum \limits_{n,n_{0}}^{}\frac{\mathcal{H}}{kT}]}
\end{equation}
The probability of one of the $2^N$ configurations defines the Gibbs factor~(\ref{05})
\begin{equation}
\label{05}
P_{i}(h,T)=\frac{e^{-\sum \limits_{n,n_{0}}^{}\frac{\mathcal{H}}{kT}}}{\sum \limits_{s_{1}}^{}
\sum \limits_{s_{2}}^{}\cdot\cdot\cdot
\sum \limits_{s_{n}}^{}\cdot\cdot\cdot
\sum \limits_{s_{N}}^{}e^{[-\sum \limits_{n,n_{0}}^{}\frac{\mathcal{H}}{kT}]}}
\end{equation}

From the statistical physics point of view, if the system $(+J)$-spins, described by the Hamiltonian (\ref{03}), located in an external magnetic field $h$, coinciding with the sign of the spin excess, at $T = 0$, then it must in the global energy minimum that corresponds to the magnetic state of complete ordering (ferromagnetism) and the probability of this event is equal to one, according to (\ref{05}), where the denominator, can leave the most important term in the sum equal to the numerator. The instantaneous change in the sign of the external field in the $-h$ should lead to an instantaneous change in the sign of the spin excess to reduce the Zeeman energy and, consequently, the transition to the symmetric configuration, because in this case, the degeneracy of the most likely (here the ground state) of the system is equal two.

At finite temperature $T\neq0$ and $T<T_{c}$  the changes the sign of the external magnetic field $h$ to $-h$, should also lead to a reversal of the spin excess. For an infinite number of spins $N$ at finite temperature, there are an unlimited number of magnetic configurations with a same excess spin and energy, i.e. with an equal probability of realization. The changes the sign of the field at $T\neq0$ and $T<T_{c}$ should increase the probability of symmetric configurations with opposite value of the spin excess. Currently available methods for Monte Carlo simulations, in particular, for example, the algorithm Metropolis-Hastings to achieve equilibrium implies that the motion of the system in state space similar to a Markov process, where the probability of each successive configurations depends on the previous implementation~(\ref{06})
\begin{equation}
\label{06}
P(E_{0})\rightarrow P(E_{1},E_{0})
\rightarrow P(E_{2},E_{1})
\rightarrow\cdot\cdot\cdot\rightarrow
P(E_{n},E_{n-1})
\end{equation}

\begin{figure}[h]
\centering{\includegraphics[width=70mm]{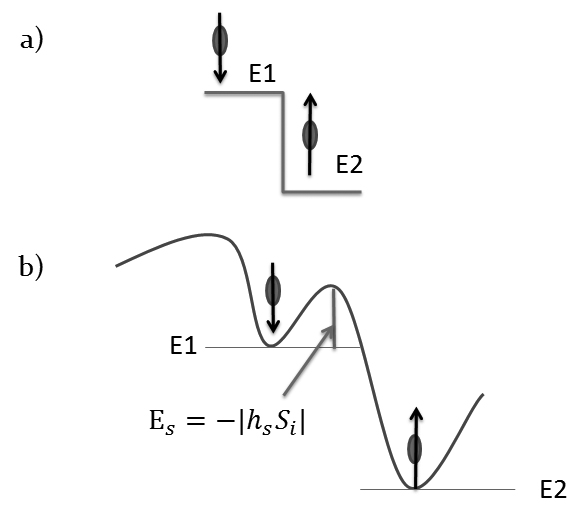}}
\caption{Variants of the transition to a state of minimum energy: a) without a field switch; b) taking into account of the field switching.}
\label{fig:4}       
\end{figure}

The movement toward equilibrium in such an approach is carried out by successive reversal individual spins in accordance with rule~(\ref{06}).
In the Ising model is used for simulation hysteresis phenomena Monte Carlo method can introduce a positive anisotropy field $h_{an}$, supporting the sign of the spin excess~(\ref{07})
\begin{equation}
\label{07}
\mathcal{H}=-J\sum \limits_{n,n_{0}}^{}S_{n}S_{n+n_{0}}-h\sum \limits_{n}^{}S_{n}-
h_{an}\Bigr|\sum \limits_{n}^{}S_{n} \Bigr|
\end{equation}
whereas treatment of the magnetization will occur with equal energy of the system of spins in the external magnetic field and the anisotropy energy.
To study the phenomenon of magnetic hysteresis in the model takes into account the spin-flip probability, which one depends on the average switching field and temperature, shown in  Fig.~\ref{fig:4}). The transition to a state corresponding to the minimum energy is possible only after overcoming a potential barrier of anisotropy.

Metropolis algorithm taking into account the critical field of switching is as follows:
\begin{enumerate}
\item We calculate the interaction energy of the spin with its neighbors in the original position $E_{1}$ and  in the new $E_{2}$ and the energy of the new configuration is compared with the energy of the old one;
\item 	The new configuration is accepted and becomes the initial for the next step, if $E_{2}<E_{1}$, otherwise the calculated probability of reversal $p$ and generated a random number from the interval $(0,1)$~(\ref{08}):
\begin{equation}\label{08}
p = \left\{
\begin{array}{rcl}
1, \ if E_{2}<E_{1} \\
e^{-\frac{\Delta E}{T}}, \ if E_{2}>E_{1}\\
\end{array}
\right.
\end{equation}

where $\Delta E = E_{2}-E_{1}$;
\item If $p$ is greater than this random number, then the new configuration is accepted, otherwise it is rejected, and the old configuration remains the initial for a new attempt;
\item In addition to the spin-flip probability shown above, take into account the spin-flip probability, calculated using the energy of the field switch, which prevents the reversal~(\ref{09}):
    \begin{equation}
\label{09}
p=e^{-\frac{\Bigr|S_{i}h_{s}\Bigr|}{T}}
\end{equation}
    and then generate a random number from the interval (0,1);
\item If the probability of revolution greater than this random number, then the MC-steps (1-3) shown above, otherwise do not change the orientation of the spin and move to the next spin.
\end{enumerate}

The critical field of switching of a magnetic particle (macrospins) - field,  which one need overcome to change the magnetization of the magnetic particles. The effective switch field is introduced to account for the spin-orbit interaction, which leads to the well-known phenomenon of anisotropy in the macroscopic scale. In the ultrafine materials can be observed scatter in the values of critical fields, the so-called coercive spectrum. In our model of epitaxial nanostructures in used approximation all the spins interact with some introduced an average effective field, which supports the direction of the spin. The reversal of the spin occurs when the equality of the Zeeman energy (energy of the spin in an external field) and the energy of the spin in the effective field switching. The probability of switching is greater, the higher the temperature, as the temperature increases the probability of thermodynamic fluctuations, the probability of overcoming the energy barrier created by the effective field and thus the probability of switching of the local energy minimum. We used the following values of the field shift for the samples with different number of monolayers: 1.5 ML - 2 r.u., 2.0 ML - 8 r.u., 2.5 ML - 4 r.u., 3.0 ML - 2 r.u. Different values of the field shift due to the need to provide qualitative agreement with experiment.

\section{Visualization of the magnetization reversal in an external magnetic field}
\label{sec:6}

\begin{figure}
\centering{\includegraphics[width=110mm]{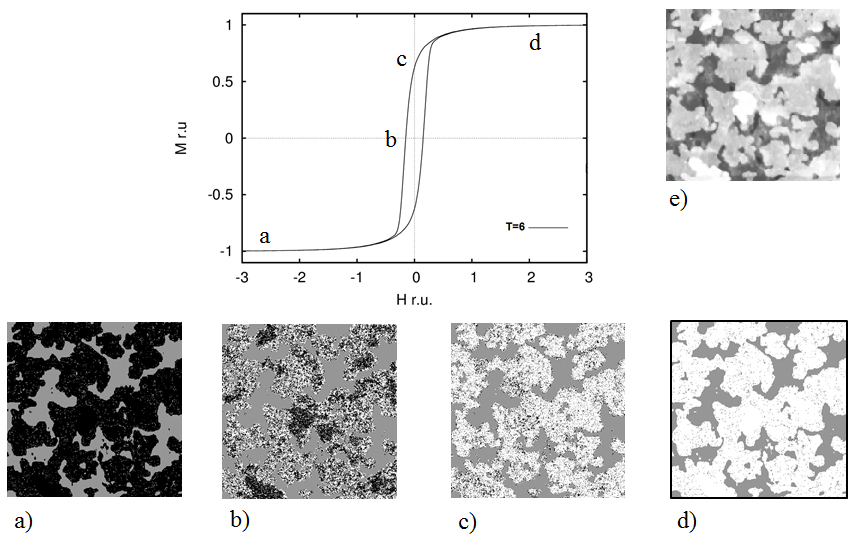}}
\caption{Hysteresis loop for a sample of 2.5 ML; a-d) Simulated PMOKE-image; e) STM-image of a sample of 2.5 ML.}
\label{fig:5}       
\end{figure}

The practical value of the proposed method also consists in the fact that the results of MC simulations of collective effects in nanoobjects can be compared with experimental results using the PMOKE - parallel magneto-optical Kerr effect, which allows directly observe the magnetic state of the surface nanostructure, as shown Fig.~\ref{fig:5}. Formation of the simulated PMOKE-image is as follows: take into account only the surface atoms of the sample, and if the spin up, then put a white pixel, the spin down - black. The approach to simulation images allow you to visually the process of magnetization reversal of the spins of the sample.
\\
\\
\section{Conclusion}
\label{sec:7}

In this paper we was presented model for processing data obtained with a scanning tunneling microscope (STM) and atomic force microscope (AFM) and the calculation of the magnetic and structural properties of the quasi-nanocluster magnets and simulation of magnetic hysteresis phenomena.

The parameters for the transition to ferromagnetic state at $T = 0$, using the critical concentration $p_{c}$, calculated. The simulation results Co-nanostructures and theoretical estimates are in qualitative agreement with experiment for determining the concentrations of phase transitions in the ferromagnetic state.

Account of the average effective field of switching allows you to simulate the phenomenon of magnetic hysteresis in the investigated epitaxial nanostructures. The developed model, the algorithm and the created on their basis software has good scalability, which is caused by the use of the scheme independent computation.

This work was supported by Ministry of Education and Science of the Russian Federation (07.514.11.4013, 02.740.11.0549, 14.740.11.0289).

\end{document}